\documentclass[12pt]{iopart}

\usepackage{graphicx}
\usepackage{epstopdf}
\usepackage{color}
\usepackage{appendix}

\usepackage{iopams}  
\begin{document}

\title[]{Dynamical crossovers in Markovian exciton transport}

\author{Hoda Hossein-Nejad$^{1,2}$, Alexandra Olaya-Castro$^2$,
Francesca Fassioli$^{2,3}$, and
Gregory  D Scholes$^3$}

\address{$^1$Department of Physics, University of Toronto, 60 St. George Street, Toronto, Ontario  M5S 1A7, Canada, }
\address{$^2$Department of Physics and Astronomy, University College London, Gower Street, London WC1E 6BT, UK} 
\address{$^3$Lash-Miller Chemical Laboratories, Institute for Optical Sciences and Centre for Quantum Information and Quantum Control, University of Toronto,  80 St. George Street, Toronto, Ontario, M5S 3H6 Canada  }

\ead{h.hossein-nejad@ucl.ac.uk, a.olaya@ucl.ac.uk}
\submitto{\NJP}
\begin{abstract}

\end{abstract}
The large deviation theory has recently been applied to open quantum systems to uncover dynamical crossovers in the space of quantum trajectories associated to Markovian evolutions. Such dynamical crossovers are characterized by qualitative changes in the fluctuations of rare quantum jump trajectories, and have been observed in the statistics of coherently driven quantum systems. In this article we investigate the counting statistics of rare quantum trajectories of an undriven system undergoing a pure relaxation process, namely, Markovian exciton transport. We find that dynamical crossovers occur in systems with a minimum of three interacting molecules, and are strongly activated by exciton delocalization and interference. Our results illustrate how quantum features of the underlying system Hamiltonian can influence the statistical properties of energy transfer processes in a non-trivial manner.

\section{Introduction}
Full counting statistics (FCS) is a powerful mathematical tool that enables the characterization of a counting process by systematic computation of the generator of its statistics. The technique was originally developed for theoretical modelling of mesoscopic charge transport \cite{Levitov:1993, Levitov:1996}, where current fluctuations can provide vital information on the nature of the transport mechanism \cite{Blanter:2000, Nazarov:2002, Bagrets:2003, Belzig:2005, Braggio:2006, Gustavsson:2006}. More generally, FCS is of importance in the theory of open quantum systems as it provides an alternative characterization of the steady-state properties of a quantum system subjected to noise \cite{Zheng:2003, Zheng2:2003, Flindt:2005}. The applicability of FCS to open quantum systems stems from the realization that the dynamics of quantum systems in noisey environments can be described in terms of evolutions of ensembles of trajectories \cite{Plenio:1998}. A trajectory, in this context, is a realization of the system's dynamics subjected to particular bath-induced jumps at stochastic intervals, and represents a sequence of configurations in the time domain. It is therefore the discrete nature of the quantum jumps that renders applicability to the tools of FCS. 
\\

In a recent contribution Garrahan and Lesanovsky applied the mathematical framework  of FCS to investigate the dynamical phases of quantum systems subjected to Markovian noise \cite{Garrahan:2010}. FCS states that a knowledge of the generating function enables complete characterization of the statistics. The contribution of Ref. \cite{Garrahan:2010} was to re-interpret the counting field of FCS as a means by which sub-ensembles of rare trajectories can be accessed, and to investigate how the statistical generator is changed as a function of the counting field. For sub-ensembles of rare trajectories the frequency of the number of jumps after a long observation time, a quantity known as the activity, deviates from the jump rate in the steady-state, i.e.: from the jump rate of the entire ensemble of trajectories. 
By unravelling the trajectory space of a master equation and identifying the rare sub-ensembles, one can uncover subtle dynamical features, such as dynamical transitions, that cannot be determined from the static properties, or the ensemble averaged dynamics \cite{Ates:2012}. Dynamical transitions occur if sub-ensembles of trajectories with different activities demonstrate qualitatively different fluctuations. They have so far been predicted in photon counting \cite{Garrahan:2010, Budini:2010, Garrahan:2011, Genway:2012}, electron counting \cite{Li:2011}, quadrature measurements \cite{Hickey:2012}, and dissipative quantum walks \cite{Garnerone:2012}, suggesting the possibility of their existence in a host of other systems. 
\\

The formalism developed by Garrahan and Lesanovsky is known as the $s$-ensemble theory. The parameter  $s$ has the same mathematical structure as the counting field of FCS; it also has the added feature that $s \neq 0$ allows access to sub-ensembles of rare trajectories. In some applications of the $s$-ensemble theory  rare trajectories are referred to as the ``non-equilibrium" trajectories to distinguish them from the typical trajectories. This terminology can lead to confusion as in the theory of open quantum systems, a non-equilibrium state is characterized by a finite rate of entropy production \cite{Breuer:2003}.  In the absence of external drives, rare trajectories, or typical ones, do not produce entropy. In the present article we are interested in dynamical features of quantum systems at thermal equilibrium and we will therefore avoid this terminology. 
\\

In this article we apply the \textit{s}-ensemble theory to investigate the statistics of rare trajectories of quantum evolutions describing excitation energy transport in a molecular complex. Assuming strong electronic interactions, and weak system-environment coupling, exciton relaxation can be described by Lindblad type evolution \cite{Breuer:2002}. The exciton relaxation process results in incoherent transfer between electronic eigenstates, reaching a steady-state where the average population of the eigenstates obeys the thermal Boltzmann distribution.  At a microscopic level, however, transfer of excitation still occurs in a dynamical manner in the steady-state. For instance for a system of two eigenstates at thermal equilibrium, detailed balance dictates that $k_{1 \rightarrow 2} P_1 = k_{2 \rightarrow 1} P_2$ where $k_{i \rightarrow j}$ is the rate of energy transfer from the state $i$ to the state $j$ and $P_i$ is the thermally-equilibrated population of the state $i$. We demonstrate that, in this system, despite the classical nature of the dynamics, the statistics of the trajectories are not, in general, Poissonian. We observe 
temperature-dependent dynamical crossovers for systems containing a minimum of three sites. Statistical  crossovers are found to be strongly influenced by exciton interference.  
\\
 
This article is organized as follows: in section \ref{sec_2} we introduce the conceptual and mathematical framework of the large deviation principle; the mathematical framework behind the $s$-ensemble theory. In section \ref{sec_3} the Markovian Lindblad master equation and the $s$-ensemble theory are introduced. In section \ref{sec_4} we apply the $s$-ensemble theory to investigate the statistics of rare trajectories in systems of two, three, four and seven sites and present numerical results that demonstrate the dynamical transitions in the statistics of trajectories.  The electronic and vibrational parameters of the Fenna-Matthews-Olson (FMO) photosynthetic complex are used in the computation of the numerical results, as the Lindblad master
equation considered in this article has been found to make a reasonable prediction of the steady-state of the complex.  
Section \ref{sec_5} summarizes the main conclusions of the article. 

\section{Large deviation principle and counting statistics}
\label{sec_2}
A counting process can be decomposed into a number of trajectories where in each trajectory the process occurs at different stochastic intervals  \cite{Esposito:2009}. In the context of quantum jumps, the counting process is the number of jumps observed after a given time $t$. Let us denote the probability of observing $K$ jumps after a time $t$ by $P_t(K)$, where $\sum_{K=0}^{\infty} P_t(K) =1$, and $\{ 0 \leq P_t (K) \leq 1\}$. For a counting process that satisfies the large deviation theory, $P_t (K)$ acquires a  large deviation form for large $t$. That is 
\begin{equation}
P_t (K) \simeq e^{-t \varphi (k)}
\end{equation}
where the function $\varphi (k)$ is known as the rate function \cite{Touchette:2009}, and $k = K/t$ is the jump rate.  For a simple (unimodal) counting process, $\varphi (k)$ has a local minima corresponding to the most probable jump rate after a long observation time. The generating function of a given probability distribution is defined to be
\begin{equation}
Z_t (s) = \sum_{K=0}^{\infty} 
P_t (K) e^{-sK}
\end{equation}
 where $s$ is a conjugate variable to $K$, and is known as the counting field in FCS. If the counting process has a large deviation form, the generating function can be written as
\begin{equation}
Z_t (s) \simeq e^{t \theta (s)}.
\end{equation}
$Z_t (s) $ is now referred to as the large deviation function, while $\theta (s)$ is the cumulant generating function. The cumulant generating function and the rate function are related via a Legendre transform, 
\begin{equation}
\theta (s) = -\min_k [\varphi (k) + ks]
\end{equation} 
The average and the standard deviation of the number of jumps in the steady-state can be deduced from the cumulant generating function, and are given by
\begin{equation}
\langle k\rangle =  -\theta'(0)
\end{equation}
\begin{equation}
\langle k^2 \rangle - \langle k \rangle^2 = 
\theta'' (0).  
\end{equation}
Similarly, higher derivatives of $\theta(s)$ evaluated at $s=0$, are the higher order statistical cumulants. Full counting statistics (or the steady-state statistics) is therefore recovered by looking at the derivatives of the cumulant generating function at $s=0$.
The applicability of the large deviation approach to the identification of dynamical phases becomes evident when the behaviour of the cumulant generating function away from $s=0$ is considered.  To illustrate how sub-ensembles of rare trajectories can be accessed via the exponential factor $e^{-sK}$,  let us construct an associated stochastic process defined by the probabilities \cite{Budini:2011}
\begin{equation}
q_t (K,s) = \frac{1}{Z_t (s)}
P_t (K) e^{-sK}
\label{6}
\end{equation}
For $s=0$ we recover the original counting process. For $s<0$ ($s>0$) the exponential factor amplifies (lowers) the probabilities of rare events.  The coupling field $s$, thus provides a systematic means by which rare events can be accessed: 
unlikely events of the counting process $\{P_t (K)\}$, correspond to typical events of the associated process  $\{q_t (K, s)\}$. Rare events can thus be accessed without generating a large number of typical events and relying on the statistics of the original counting process. 

In statistical mechanics the counting process 
is the number of microstate configurations accessible to the system under certain constraints, making 
$Z_t(s)$ the partition function. Analogously, for an open quantum system, $Z_t(s)$ is a measure of the number of trajectories that can participate in a given (rare) sub-ensemble. One can also identify $\varphi(k)$ as the entropy:  extrema of $\varphi(k)$ therefore correspond to the typical trajectories. The function $\theta (s)$ is the equivalent of a free energy and its derivative with respect to $s$, is a dynamical order parameter known as the \textit{activity}. Non-analyticities in free energy lead to abrupt jumps in the order parameter. These jumps correspond to 
first-order thermodynamic phase transitions, or (first-order) dynamical transitions in the space of trajectories of a quantum system.  

First-order dynamical transitions at $s=0$ have important consequences for the steady-state dynamics, and indicate the coexistance of dynamical phases, characterized by distinct activities.The implications of transitions at $s \neq 0$  are not always clear and tend to be problem-specific. Sometimes the original master equation can be mapped into another master equation, with different jump operators, that undergoes dynamical transitions at $s = 0$ \cite{Garrahan:2010}. In the present article we endeavour to explore the possibility of dynamical transitions in electronic energy transfer (EET) as the size of the system is varied, and will not explore such mappings. 

\section{$s$-ensemble theory and electronic energy transfer}
\label{sec_3}
In this section we apply the large deviation theory 
to investigate the trajectories of a particular open quantum system, modelled by the Markovian Lindblad master equation. The system of interest consists of a number of molecules (sites) coupled via electronic couplings and weakly interacting with their environment. Each molecule is modelled as a two-level system in its electronic coordinate, where the two levels correspond to the ground and the first excited electronic states. 
The electronic degrees of freedom are assumed to be diagonally coupled to a bosonic bath of harmonic oscillators. 
We assume that there is one excitation in the system that can be coherently shared between multiple molecules (a molecular exciton). The Hamiltonian of this system is given by
\begin{equation} 
 H = -\sum_{ij} J_{ij} |i \rangle \langle j |
+ \sum_{i=1}^{2} |i \rangle \langle i |  \Big[  \epsilon
 + \frac{1}{2} \sum_k h_{ik} (b_k^{\dagger} + b_k) \Big]
 +  \sum_k \omega_k b_k^{\dagger}b_k 
 \label{dimer_ham}
 \end{equation}
 where $J_{ij}$ is  the electronic coupling between the ground state of site $i$ and excited state of site $j$, the ket $|i\rangle$ represents the state in which site $i$ is excited and all other sites are in the ground state, $h_{ik}$ is the electron-phonon coupling between site $i$ and mode $k$ of the bath, $\{b_k^{\dagger}, b_k \}$ are the raising and lowering operators for mode $k$ of the bath, and $\omega_k$ are the bath frequencies. 

Due to the assumption of weak system-bath interaction, the exciton states are delocalized and the influence of the environment is to induce jumps between them at stochastic intervals. 
In the steady-state, the average population of the exciton states is given by the Boltzmann factor: $P_i = e^{-\beta \omega_i}/Z$, where $P_i$ is the population of the exciton state with energy $\omega_i$ and $Z$ is the partition function. From a microscopic standpoint, dynamic transfer of excitation still persists in the steady-state. This microscopic transfer consists of incoherent jumps between the excitonic eigenstates. For instance for a system of two sites, steady-state implies that $k_{1 \rightarrow 2} P_1 = k_{2 \rightarrow 1} P_2$, where $k_{i \rightarrow j}$ is the transfer rate from the eigenstate $i$ to the eigenstate $j$. Moreover, the ratio of the transfer rates satisfies  detailed balance: $k_{1 \rightarrow 2} = k_{2 \rightarrow1} e^{-\beta \omega_{12} }$, where $\omega_{12}$ is the energy difference between the two states. $k_{1 \rightarrow 2} P_1 $ is also the average number of jumps per second between the two states: upward and downward jumps occur with equal frequency  in the steady-state.  For rare sub-ensembles of trajectories, the frequency of the jumps deviates from the equilibrium rate $k_{1 \rightarrow 2} P_1 $.

To compute the statistics of jumps in the steady-state, one can compute the cumulant generating function of this process directly from the Lindblad equation. 
 The Lindblad equation as applied to electronic energy transfer,
 for weak system-bath coupling with
 secular approximation is given
 by \cite{Breuer:2002}
\begin{equation}
 \dot{\sigma} (t)= 
-i[H_s, \sigma (t)] 
+ 
\sum_{\omega,m }
\gamma (\omega)
\bigg[
A_m (\omega) \sigma (t) A_m^{\dagger} (\omega)
- \frac{1}{2}
 \{
 A_m^{\dagger} A_m(\omega), \sigma (t)
  \}
\bigg]
\label{eqn9}
\end{equation}
where the commutator describes the unitary evolution and the second term describes relaxation and dephasing in the eigenstates
 $\{ | \alpha \rangle \}$ of the unperturbed system Hamiltonian $H_s$ (neglecting the Lamb-shift Hamiltonian). The Lindblad operators are given by
 $A_m (\omega) = \sum_{\omega = \epsilon_{\alpha} - \epsilon_{\alpha'} } 
 c_m^{*} (\alpha) c_m(\alpha ') | \alpha \rangle \langle \alpha' |$ where $\{\omega\}$ is the energy difference associated with single excitation eigenstates, and $c_m^{*} (\alpha) c_m(\alpha')$ is the \textit{exciton interference} at site $m$. The eigenstates $\{ |\alpha \rangle\}$ are delocalized over the sites, with amplitude $c_m (\alpha)$ at site $m$, that is
  \begin{equation}
 |\alpha \rangle = \sum_m c_m(\alpha) |m\rangle
 \end{equation}
where $\{|m\rangle \}$ are the localized wavefunctions. For two eigenstates $|\alpha \rangle$ and $| \alpha'\rangle$, we define the `intensity factor' to be
 \begin{equation}
 I(\alpha, \alpha')= \sum_m |c_m (\alpha') |^2 |c_m(\alpha)|^2.
 \end{equation}
 The intensity factor adds up the squared magnitude of the exciton interference at all sites and determines the contribution of the electronic Hamiltonian to the transfer rate between two exciton states. 
 The factor  $\gamma(\omega)$ in Eq.~(\ref{eqn9}) determines the bath contribution to the exciton transfer rate and 
 is related to
  the spectral density of the bath $J(\omega)$, via the expression
  $\gamma (\omega) = 2 \pi J(|\omega|) |n(\omega)|$, 
  where $n(\omega) = (e^{\beta \omega } - 1)^{-1}$ is the thermal
  occupation number. 
  In the present treatment it is assumed that each molecule interacts  with an identical and independent phonon bath. 
The dissipative term in the Lindblad equation describes the incoherent jumps between the exciton
states as induced by the phonon bath. In the quantum jump approach 
the Hamiltonian evolution of the quantum system is
interrupted by jumps at stochastic intervals.  A time-record of deterministic (non-unitary) evolution of the wavefunction interrupted by stochastic jumps constitutes a trajectory of the system. An averaging over a large number of such trajectories recovers the reduced density matrix  $\sigma(t)$ \cite{Plenio:1998}.

If $K$ jumps are observed after a time $t$, 
the (unnormalized) reduced density matrix of the system prior to detection is given by 
 the projection of the total density matrix onto the subspace of 
$K$ events \cite{Zheng:2003, Zheng2:2003}. We denote this projected density matrix by $\sigma^{(K)} (t)$. 
 Conversely, the total density matrix can be written as 
an expansion of all projected density matrices, that is
$\sigma (t) = \sum_{K=0}^{\infty} \sigma^{(K)} (t)$. The probability of observing 
$K$ events after a time $t$ is given by $P_t (K)=$Tr$[\sigma^{(K)} (t)]$.
At this point
for the sake of clarity we  
restrict the analysis
to the statistics of jumps between two exciton states separated 
in energy by $\omega_1$. Note that the total system may still contain an arbitrary number of exciton states. 
It 
can be   
shown that the Laplace transform of the projected density operator, defined by
the expression
 $\sigma_s (t) = \sum_0^{\infty} \sigma^{(K)} e^{-sK}$ obeys the equation $ \dot{\sigma_s } (t) = 
 W_s [\sigma_s]$, where the superoperator $W_s$ is defined to be
 \begin{eqnarray}
&  W_s [\sigma_s (t) ]  = 
-i[H_s, \sigma_s (t)] 
+ \sum_{m}
\gamma (\omega_1)
e^{-s}
A_m (\omega_1) \sigma_s (t) A_m^{\dagger} (\omega_1)
\label{above_eqn15}
&\\&
+ 
\sum_{m}
\bigg[ 
\sum_{\omega \neq \omega_1} \gamma (\omega)
A_m (\omega) \sigma_s (t) A_m^{\dagger} (\omega)
- \frac{1}{2}
\sum_{\omega} \gamma (\omega)
 \{
 A_m^{\dagger} (\omega)A_m(\omega), \sigma_s (t)
  \}
\bigg]
\nonumber
\end{eqnarray}
Note that $\sigma_s (t)$ has the same structure as the variable $q_t (K, s)$ in Eq.~(\ref{6}). Typical trajectories of $\sigma_s (t)$ thus correspond to rare trajectories of the original master equation. The ensemble of trajectories generated by Eq.~(\ref{above_eqn15}) is known as the $s$-ensemble \cite{Garrahan:2010}.  
Physical time evolution is retrieved at $s= 0$, whereas $s \neq 0$ contains information with regards to rare
 trajectories of the system. $s<0$ ($s>0$ ) generates the active (inactive) trajectories where the number of jumps are significantly higher (lower) than average.   
 
The Markovian nature of the Lindblad equation enables us to 
compute the statistical 
moments of the number of jumps in the steady-state via the large deviation principle. Large deviation theory states that for a Markov process, the largest real
 eigenvalue of the superoperator $W_s$ is the cumulant generating
 function of the statistics  \cite{Touchette:2009}. Following the notation of the previous section, we denote this generator by $\theta (s)$. 
In the next section we compute the cumulant generating function and the associated statistical moments for a system of two, three, four and seven sites.

\section{Results and discussion}
\label{sec_4}
\begin{figure}[htbp]
\vspace{0.1in}
\center
\includegraphics[scale=0.23]{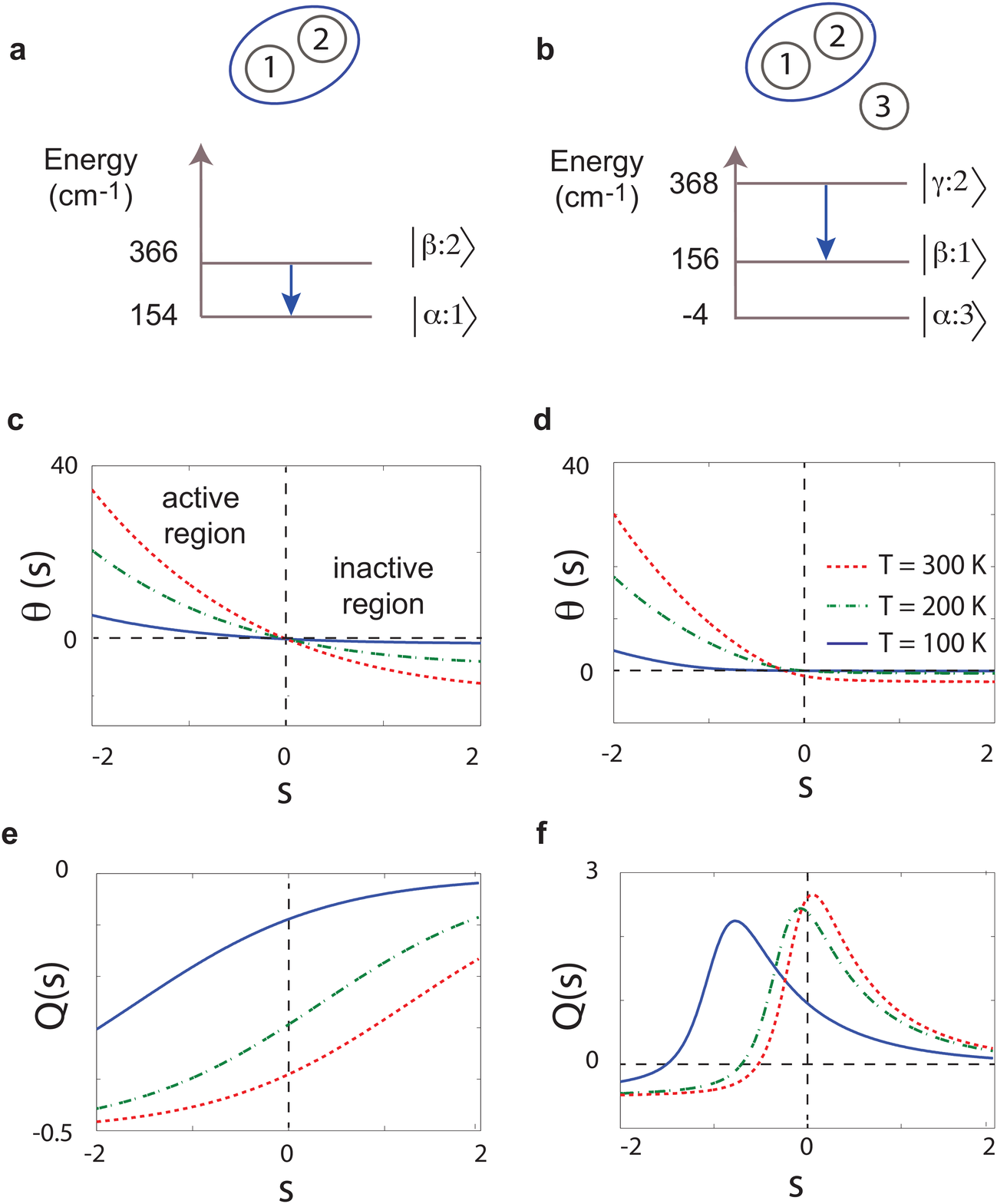}
\caption[Cumulant generating function for a system of two and three sites.] {a) Excition states with only two sites present and the corresponding schematics of the interacting sites. The circle containing both sites indicates that they are strongly interacting.  b)
Excition states in the three site model. The three site model consists of a strongly interacting pair,  weakly interacting with a third site. 
 For each case the statistics of jumps between the excitons associated with sites 1 and 2 are computed. c) $\theta (s)$ for the two site model, d) and the three site model.  e) $Q(s)$ for the two site,  f) and the three site model.  The site energies are $\epsilon_1 = 200$~cm$^{-1}$, $\epsilon_2 = 320$~cm$^{-1}$, $\epsilon_3 = 0$~cm$^{-1}$ and the electronic couplings are $J_{12} = -87.7$~cm$^{-1}$, $J_{13} = 5.5$~cm$^{-1}$, $J_{23} =  30.8$~cm$^{-1}$ \cite{Adolphs:2006}. }
\label{fig_1}
\end{figure}

Adopting the electronic and vibrational parameters 
of the FMO complex, we first consider two and three sites (chromophores)  and study the statistics of trajectories that contain jumps between two given exciton states.  
  We subsequently include further sites in the system and study the statistics of selected jumps.
We describe the coupling of the electronic system to the environment via the spectral density \cite{Ishizaki:2009}, 
 \begin{equation}
 J(\omega) = \frac{2 E_r}{\pi} \frac{\omega \omega_c}{\omega^2 + \omega_c^2}
 \end{equation}
where $E_r = 35$~cm$^{-1}$ is the reorganization energy $E_r = \int_0^{\infty} \frac{J(\omega)}{ \omega } d \omega$,
and $\omega_c = 150$~cm$^{-1}$ is the cutoff 
frequency of the bath \cite{Mohseni:2008, Adolphs:2006}. This form of the spectral density describes the low frequency contribution of the environmental fluctuation spectra correctly and captures the main features of the dynamics. The sensitivity of the dynamical transitions on the form of the spectral density is an interesting topic that will not be addressed in the present work. 
We compute the matrix $W_s$ 
and solve for the cumulant generating function $\theta (s)$. 
Figure \ref{fig_1} (a) and 1(b)
 illustrate the exciton states where two and three sites are included
 in the dynamics respectively. The two site model includes  sites 1 and 2, and the three site model includes  sites 1, 2 and 3. The site indices are the standard indices assigned to the chromophores in the FMO complex \cite{Fassioli2:2010}.
 The exciton states are labelled via a Greek letter and a number, where the Greek letter represents the position of the exciton state in the energy ladder, and the number indicates the site that carries the maximum exciton amplitude. For instance $| \alpha:1\rangle$ represents the lowest eigenstate, and indicates that this eigenstate is predominately localized on site 1. Due to significant exciton localization in the FMO protein, each exciton can be unambiguously assigned to a unique site index. 

  Figure \ref{fig_1} (c) and (d)
 illustrate the cumulant generating function versus $s$ for different temperatures.  The plots suggest that the 
  two cases have similar statistics in the active phase, but begin to
  diverge close to $s=0$. Figure \ref{fig_1} (e)  illustrates the Mandel parameter for the case of two sites.   Mandel parameter is an indicator of the nature of first-order statistical correlations in the system and is defined to be 
\begin{equation}
 Q(s) = \frac {\langle k^2 \rangle - \langle k \rangle^2 }{\langle k \rangle} -1 = - \frac{\theta'' (s)}{\theta' (s)} - 1.
 \end{equation}
$Q = 0$ implies Poissonian statistics where probabilities of subsequent events are uncorrelated. $Q > 0$ ($Q<0$) indicates super-Poissonian (sub-Poissonian) statistics where subsequent jumps exhibit first-order correlations. Strictly, a dynamical transition occurs when $\theta (s)$ exhibits non-analytic  behaviour. Such non-analyticities are not observed in our system of interest due to its finite size. Instead, we observe smooth dynamical crossovers characterized by a change in the sign of the Mandel parameter.  In the vicinity of the crossover point, one can also identify a local maximum in the Mandel parameter. The maximum is indicative of  coexistance of two dynamical phases characterized by distinct activities. In particular, if a maximum occurs at $s=0$, there is coexistance of dynamical phases in the steady-state.

For the case of two sites, the Mandel parameter is negative for all $s$.  This indicates sub-Poissonian correlations (anti-bunching) where a waiting period for re-excitation 
is required before the system can relax back to the lower state. As  $s$ is increased, the Mandel parameter approaches zero (Poissonian limit). In this limit there are so few jumps that correlations vanish asymptotically. 
\\

It is striking that anti-bunching is observed even though the transitions are thermally activated, and the steady-state dynamics obey a classical rate equation.  The classical nature of the correlations can be confirmed by considering the rate equation characterized by the following dynamical matrix
\[
W_{class}(s) =
\left[ {\begin{array}{cc}
-\kappa   & ~~~~~~~ \Gamma e^{-s}  \\
~~\kappa   &  -\Gamma  \\
 \end{array} } \right]
\]
where $\kappa$ and $\Gamma$ are the relevant equilibrium transfer rates, and $\Gamma = \kappa  e^{- \beta \omega}$ as required by detailed balance.   By diagonalizing $W_{class}(s)$, we arrive at the following expression for $Q(s)$
\begin{equation}
Q(s) = \frac{-2 \kappa \Gamma}{(\kappa + \Gamma)^2 - 4 \kappa \Gamma  (1 - e^{-s})}
e^{-s} <0  ~~~~~  \forall s
\end{equation}
Indeed, the same expression for $Q(s)$ is obtained if the full Lindblad equation with dephasing terms is instead considered. 
\\

The corresponding plot of the Mandel parameter for a system of three sites is shown in Fig. \ref{fig_1} (f).  The trajectories exhibit a dynamical crossover in their statistics, i.e.: the correlations change from sub-Poissonian ($Q(s)<0$) to super-Poissonian ($Q(s)>0$) as $s$ is varied.  
The crossover point is temperature dependent and is closer to $s=0$ at higher temperatures. 

One can continue to add more sites to the model and investigate the statistics of a particular jump, or the collective statistics of a number of jumps. For instance the collective statistics of all jumps to the lowest eigenstate can be investigated. 
In general, beyond a few sites there are many possible jumps and there is no obvious advantageous collective counting strategy. The statistics of selected transitions for the four site and the seven site models, are explored in the remaining part of this section.
\\

\begin{figure}[htbp]
\vspace{0.1in}
\center
\includegraphics[scale=0.23]{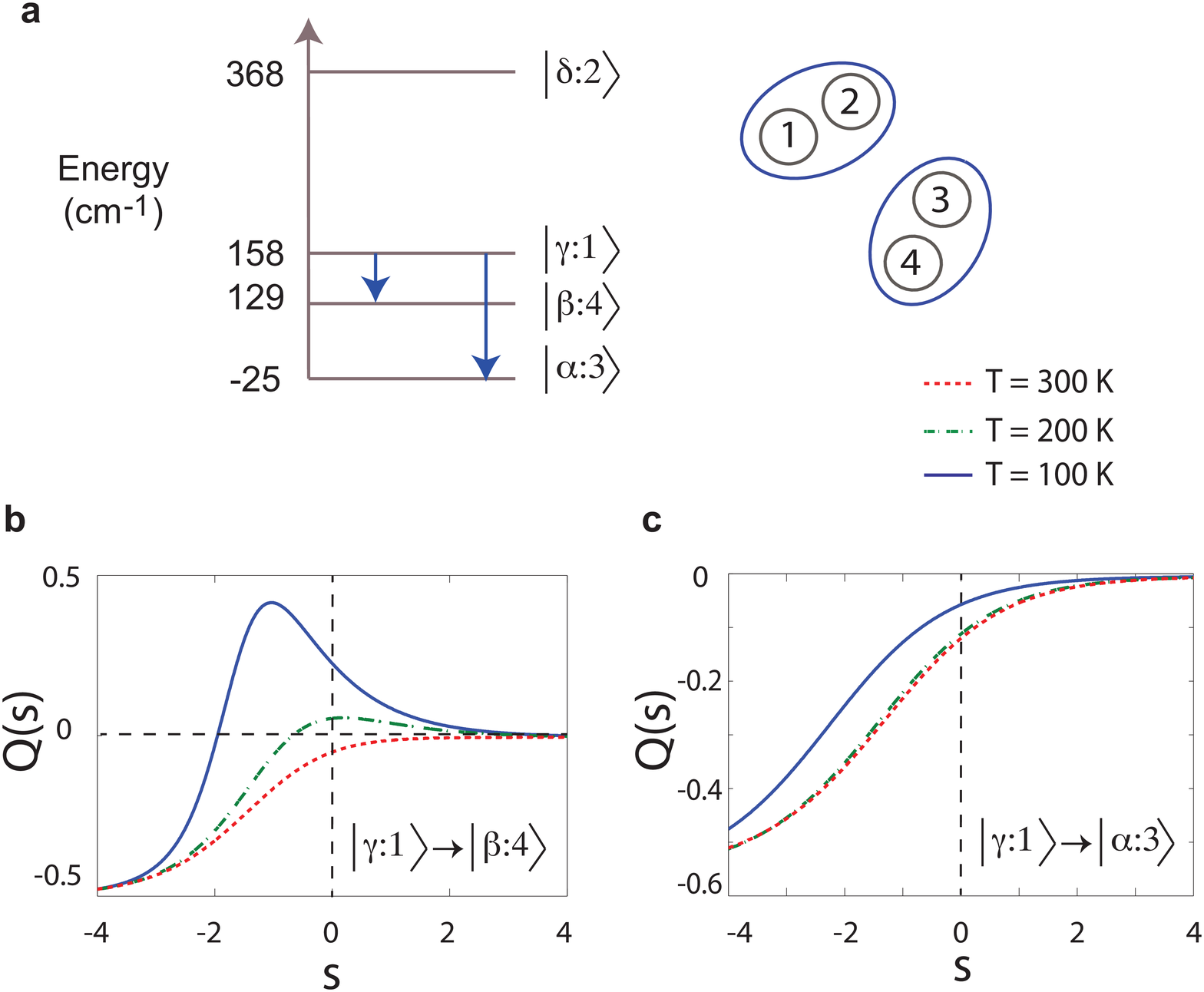}
\caption[]{a) Exciton states for the four site model, and the corresponding schematics of the interacting sites. The four site model consist of two groups of strongly interacting dimers. b) and c) Mandel parameter versus $s$ for selected downward 
jumps in the four site model. The additional site energy is $\epsilon_4 = 110$~cm$^{-1}$, and the additional electronic couplings are $J_{14} = -5.9$~cm$^{-1}$, $J_{24} = 8.2$~cm$^{-1}$, $J_{34} = -53.5$~cm$^{-1}$ \cite{Adolphs:2006}. }
\label{fig_2}
\end{figure}

Four interacting sites exhibit the combination of the features of the two site and the three site models, as illustrated in Fig. \ref{fig_2}. Figure \ref{fig_2} (b) exhibits the new feature that typical trajectories ($s=0$) have different statistics at different temperatures: anti-bunching at higher temperatures and bunching at lower temperatures.  In Fig. \ref{fig_2} (c) the trajectories remain anti-bunched over the computed temperature range, similar to the results of the two site model. 

 We consider all possible pairwise jumps of the four site model and conclude that 
for sufficiently small $s$ all trajectories become anti-bunched and approach a temperature independent but transition dependent value of the Mandel parameter. For sufficiently large $s$, on the other hand, all trajectories approach the Poissonian limit. Trajectories that undergo a statistical crossover, approach the Poissonian limit from the positive values of $Q(s)$;  those that do not, remain in the $Q(s)<0$ portion. Approach to the Poissonian limit from the $Q(s)>0$ side, is therefore a sufficient indicator of the existence of statistical crossovers. These conclusions are also observed to be valid for larger systems.

\begin{figure}[htbp]
\vspace{0.1in}
\center
\includegraphics[scale=0.23]{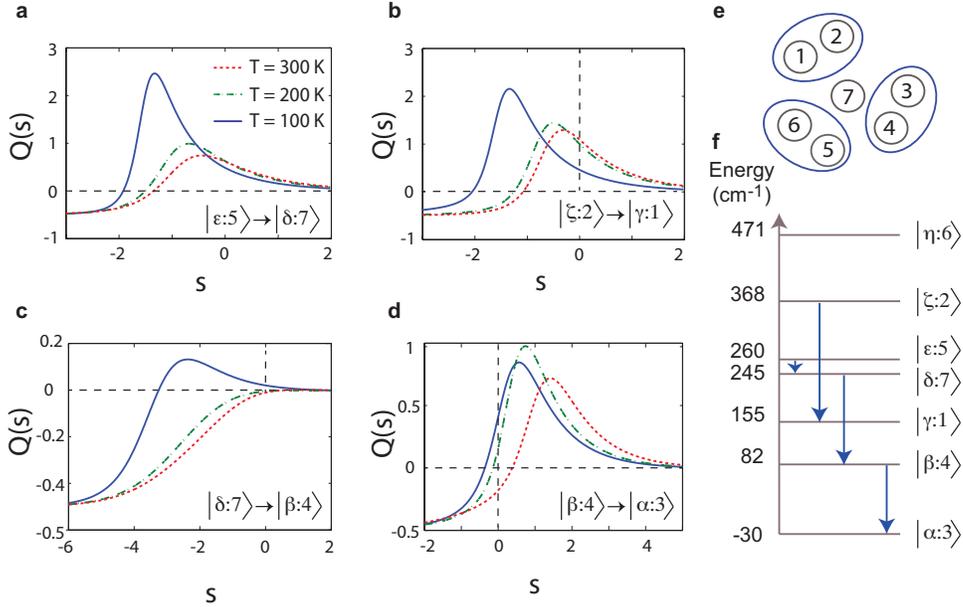}
\caption{a) to d) Mandel parameter for selected downward transitions. e) The seven site system consists of three strongly interacting dimers, coupled weakly to a seventh site. f) Exciton states of the seven site system and the selected transitions. }
\label{fig_3}
\end{figure}

Finally, statistics of selected jumps from the seven site system of the FMO complex are illustrated in Fig. \ref{fig_3}. 
All plots reveal statistical crossovers at certain temperatures. 
The crossovers are most likely to occur in the unravelling of the jumps associated to fastest exciton transfer rates. The transfer rate between two exciton states separated in energy by $\omega$ is given by $\Gamma (\omega) = \gamma (\omega) \sum_m |c_m (\alpha)|^2 |c_m (\alpha ')|^2$. In this system, the relative magnitude of the transfer rates tends to be dominated by the intensity factor $ \sum_m |c_m (\alpha)|^2 |c_m (\alpha ')|^2$, as the variation of $\gamma (\omega)$ over the energy range of interest is small. In other words, \textit{dynamical crossovers are observed in trajectories accounting for jumps between eigenstates producing the largest intensity factor}. The statistics of the rare trajectories are therefore sensitively dependent on the distribution of the electronic couplings and the degree of delocalization and interference. For exciton states that are close in energy, $\gamma (\omega)$ is large, resulting in fast transfer. Statistical crossovers are therefore also observed in this regime, despite the small intensity factor. Example of this behaviour can be seen in Fig. \ref {fig_3} (a). 

 We conclude by noting that the importance of the Markovian assumption in the present treatment poses a difficulty for many systems exhibiting EET whose environmental interactions are strongly non-Markovian. The formalism can, however, be applied to model certain types of non-Markovianity by extending the Hilbert space of the system Hamiltonian to contain the environmental modes that capture the strong non-Markovian features \cite{Mazzola:2009, Breuer:2004, Emary:2011}.

\section{Experimental plausibility }

Unlike the familiar problems of photon counting in atomic systems, the transitions discussed in this article are not accompanied by absorption and emission of real photons. 
This means that unless the energy exchange with the bath is somehow monitored, the trajectories cannot be 
directly probed. 
Monitoring the energetic fluctuations of the bath is certainly not a straightforward task, there is thus an experimental difficulty associated with the \textit{direct} detection of the dynamical crossovers discussed in this article. 

However, it might be possible to devise an experimental scheme that aims to identify \textit{indirect} experimental signatures of dynamical crossovers.  To illustrate this idea consider an external parameter $F$, that influences the activity of the trajectories and can be varied at will in an experiment (e.g.: temperature of the bath, separation between the sites, etc). The task is then to measure the transfer rates between all exciton pairs as this external parameter is varied. From these measurements one can construct $Q(F)$ at $s=0$, from the corresponding analytic expression for $Q(s=0)$. If $Q(F)$ has a local maximum at a given $F$, we anticipate dynamical phase coexistence in the vicinity of this point. In other words, maxima of $Q$ as a function of an external parameter are likely to indicate the mixing of two dynamical phases \cite{Ates:2012}, and thus the existence of dynamical crossovers. 
This connection can be theoretically verified by computing the probability distribution of the activity order parameter as discussed in Ref.  \cite{Ates:2012}.

\newpage

\section{Conclusion}
 \label{sec_5}
Markovian master equations are an established method of modelling relaxation dynamics of many systems exhibiting transport properties. These equations are primarily used to compute the ensemble averages, but in fact contain more information pertaining to higher order statistical moments that cannot be gleaned unless an unravelling formalism is first implemented.  By investigating the properties of the statistical generator as a function of the parameter $s$, the $s$-ensemble theory extends the unravelling formalism to sub-ensembles of rare trajectories of a quantum  process. 
This  enables the identification of dynamical phases, characterized by distinct statistical features. Such an unravelling formalism has been applied in the present article to investigate the statistics of Markovian excitation transport.
 
The steady-state of a system undergoing electronic energy transfer exhibits surprisingly rich statistics: The features that are often associated with quantum correlations such  as anti-bunching can be observed in the steady-state statistics of the thermally-activated quantum jump trajectories. 
  System size also plays a key role in the statistics of the dynamical phases as systems of two and three sites reveal qualitatively different correlations: while in the former the jumps are always sub-Poissonian, in the latter they can undergo statistical crossovers.  Transitions in larger systems show a combination of these two characteristic behaviours. Importantly, statistical crossovers are found to be activated by electronic interferences: 
 trajectories that account for jumps between two strongly interfering eigenstates (as defined in section 3) are most likely to exhibit statistical crossovers.
The structure of the spectral density can also influence the statistics by promoting electronic transitions of particular frequencies. 

 The physical implications of dynamical transitions away from $s \neq 0$ are not always clear and is the subject of debate. For instance, it has been shown that by rescaling time, it is possible to map the dynamics to another system, such that the rare trajectories of the original system are the typical trajectories of the new system  \cite{Garrahan:2010}. The new system then exhibits dynamical transitions under physical time evolution. It would be interesting to explore whether such mappings can be obtained for the present system. 
 
We leave the matter of the potential implications of the observed statistical transitions to excitation transport as an intriguing open question. As spectroscopy is the established tool of probing EET, one is faced with the question: Does a statistical crossover at s=0 leave its signature on any of the spectral features of a molecular complex? This question is the subject of our current investigation.

 \section{Acknowledgments}
 J. P. Garrahan is gratefully acknowledged for numerous stimulating discussions. C. Emary is thanked for a critical reading of the manuscript. H. H.-N. also acknowledges 
 the hospitality of University College London. 
This research was partially funded by the Engineering and Physical Research Council (EPSRC) of the UK under the grant EP/G005222/1. The Natural Sciences and Engineering Research Council of Canada (NSERC) is also thanked for financial support.  
 
 \section*{References}
\addcontentsline{toc}{section}{References}

\bibliographystyle{unsrt.bst}	
	
\bibliography{Hossein_Nejad}	

\end{document}